\documentclass[%
 reprint,
 amsmath,amssymb,
]{revtex4-2}
\setcitestyle{super}
\usepackage{graphicx}
\usepackage{dcolumn}
\usepackage{bm}
\usepackage{chemformula}
\usepackage{longtable}
\usepackage{soul}
\usepackage{xcolor}
\usepackage{float}
\usepackage{makecell}

\usepackage[colorlinks=true, linktocpage=true]{hyperref}

\begin{document}


\title{Deep learning approach to genome of two-dimensional materials with flat electronic bands}
\author{A. Bhattacharya$^{1,2}$}
\email{anupam.bhattacharya@mech.iitd.ac.in}

\author{I. Timokhin$^1$}
\author{R. Chatterjee$^3$}

\author{Q. Yang$^{1}$}
\email{qian.yang@manchester.ac.uk}

\author{A. Mishchenko$^{1}$}
\email{artem.mishchenko@gmail.com}

\affiliation{$^1$Department of Physics and Astronomy, University of Manchester, Manchester, UK}
\affiliation{$^2$Department of Mechanical Engineering, Indian Institute of Technology Delhi, New Delhi, India}%
\affiliation{$^3$Department of Physics, Indian Institute of Technology Delhi, New Delhi, India}%

\begin{abstract}
Many-body electron-electron correlations play central role in condensed matter physics; they govern a wide range of phenomena, stretching from superconductivity to magnetism, and are behind numerous technological applications. Two-dimensional (2D) materials with flat electronic bands provide a natural playground to explore this rich interaction-driven physics, thanks to their highly localized electrons. The search for novel 2D flat band materials has since attracted intensive efforts, especially now with the development of open science databases encompassing thousands of materials with computed electronic bands. In this work, we automate the otherwise daunting task of materials search and classification by combining supervised and unsupervised machine learning algorithms. To this end, a convolutional neural network was employed to identify 2D flat band materials, which were then subjected to symmetry-based analysis using a bilayer unsupervised learning algorithm. Such hybrid approach of exploring materials databases allowed us to reveal completely new material classes outside the known flat band paradigms, with high efficiency and accuracy. Our results present a genome of 2D materials hosting flat electronic bands that can be further explored to enrich the physics of electron-electron interactions.
\end{abstract}

\maketitle

Electrons in some materials have dispersionless spectrum, i.e. their energy $E_{\mathbf{k}}$ is independent of momentum $\mathbf{k}$, resulting in the formation of flat band, $E_{\mathbf{k}} \approx \mathrm{const}$. Vanishing group velocity, $\nabla_{\mathbf{k}} E \approx 0$, suppresses kinetic energy, making flat bands conducive to electron-electron interactions\cite{regnault2022catalogue}. Flat bands originate from spatial localization of wavefunctions in, for instance,  $f$-orbitals of lanthanides and actinides, lattice sites of atomic insulators, line- and split-graphs of bipartite lattices, or twisted bilayer graphene superlattices\cite{ma2020spin, Bistritzer2011}, where controlled engineering of the flat band has been demonstrated. Flat bands often exhibit topological nontriviality, resulting in a plethora of exotic physics like unconventional superconductivity in twisted bilayer graphene \cite{Cao2018a}, quantum anomalous Hall effect\cite{Li2021}, anomalous Landau levels\cite{Rhim2020}, strongly correlated Chern insulators \cite{Choi2021}, Wigner crystallization \cite{Li2021a}, unusual ferromagnetism \cite{Wang2022}, chiral plasmons\cite{Huang2022}, or bulk photovoltaic effect\cite{Ma2022}, and the list goes on. Most of these were reported in two-dimensional (2D) materials and have triggered intense search for new 2D materials hosting flat bands and novel physics. Some have focused on theoretical 2D lattices hosting dispersionless bands, like Lieb, kagome, or dice\cite{leykam2018artificial}. Meanwhile, progress in open materials science platforms, such as Materials Project\cite{jain2013commentary} and Aflow \cite{curtarolo2012aflow}, provides thousands of possible exfoliable 2D candidates from their 3D counterparts \cite{ashton2017topology, Mounet2018, Boland2021}, forming large databases made solely of 2D materials, like C2DB \cite{haastrup2018computational}, 2D materials encyclopedia \cite{zhou20192dmatpedia}, and MC2D \cite{Mounet2018}. 2D lattices hosting flat bands, mostly theoretically, can then be identified for further investigation. Alternatively, a symmetry-based approach resorting to known flat dispersions in line- and split-graphs of bipartite lattices could also help to predict new 2D lattices\cite{cualuguaru2021general,regnault2022catalogue}. However, with a limited number of known 2D flat band materials, current symmetry-based studies lack the statistical significance to make accurate predictions.

In this work, we use a combination of these two approaches - first, identify 2D flat band materials in the vast database using the convolutional neural network deep learning approach; the identified lattices are then classified based on their structural fingerprints, employing symmetry-based clustering. Our deep learning algorithm surveys the entire database with high throughput and precision. The resulting flat band sublattices are presented as clustering charts, serving as a roadmap for 2D flat band lattices in the future. Our work, therefore, offers a comprehensive and efficient path towards 2D flat band materials and, more importantly, the exciting physics they enable.

The search of flat dispersions in 2D materials has been attempted, where band flatness was defined by an arbitrary fixed bandwidth \cite{PhysRevMaterials.5.084203, duan2022inventory}. However, the identification of flat bands is not a straightforward task, even for simple band structures. When an electronic band is parametrized, the band index is determined by the order in which it appears in the energy scale. However, crossings between bands can change the band index, which could lead to large bandwidth even when a flat band exists. Hence, using parametrized bands and predefined bandwidth to identify flat bands may largely underestimate their number, therefore, risk overlooking potential flat band materials. To account for this, we put forward an unusual yet more inclusive approach, utilizing the band structure images from a database rather than the parametrized bands themselves. Current databases still lack the complete description of wavefunctions, which is a prerequisite for systematic classification of the electronic bands using machine learning algorithm based on, for example, topology, or other features\cite{Scheurer2020, Nunez2018, Nikolaj2022,kuroda2022machine}. However, the easily accessible band structure images offer a timely alternative to help identifying flat features in the bands, allowing us to harvest these 2D databases already at such an early stage while more development in data science to enrich their contents is on the way.

Our deep-learning-assisted framework enables high-throughput identification of flat bands from any database with images of band structures. Here we choose 2D Materials Encyclopedia (2Dmatpedia), currently the largest open 2D materials database, as the source of the band structure images. A convolutional neural network (CNN) was trained and applied to identify a genome of flat band materials across 2Dmatpedia. Afterwards, symmetry-based analysis using unsupervised machine learning algorithm was employed to classify flat-band materials into clusters, based on the structural fingerprints of identified corresponding sublattices. This classification framework helps to retain subtle connections among structural nuances, enabling the identification of hierarchical classes within the flat band 2D materials genome. Perhaps more importantly, it allows the prediction of entirely new flat band 2D materials outside the known paradigm, quickly expanding our 2D flat band materials toolbox. Our work, therefore, presents a new path towards the exploration of the rich and exciting interdisciplinary field across data science and physics.

\section*{Results}

\begin{figure*}
    \centering
    \includegraphics[width=\textwidth]{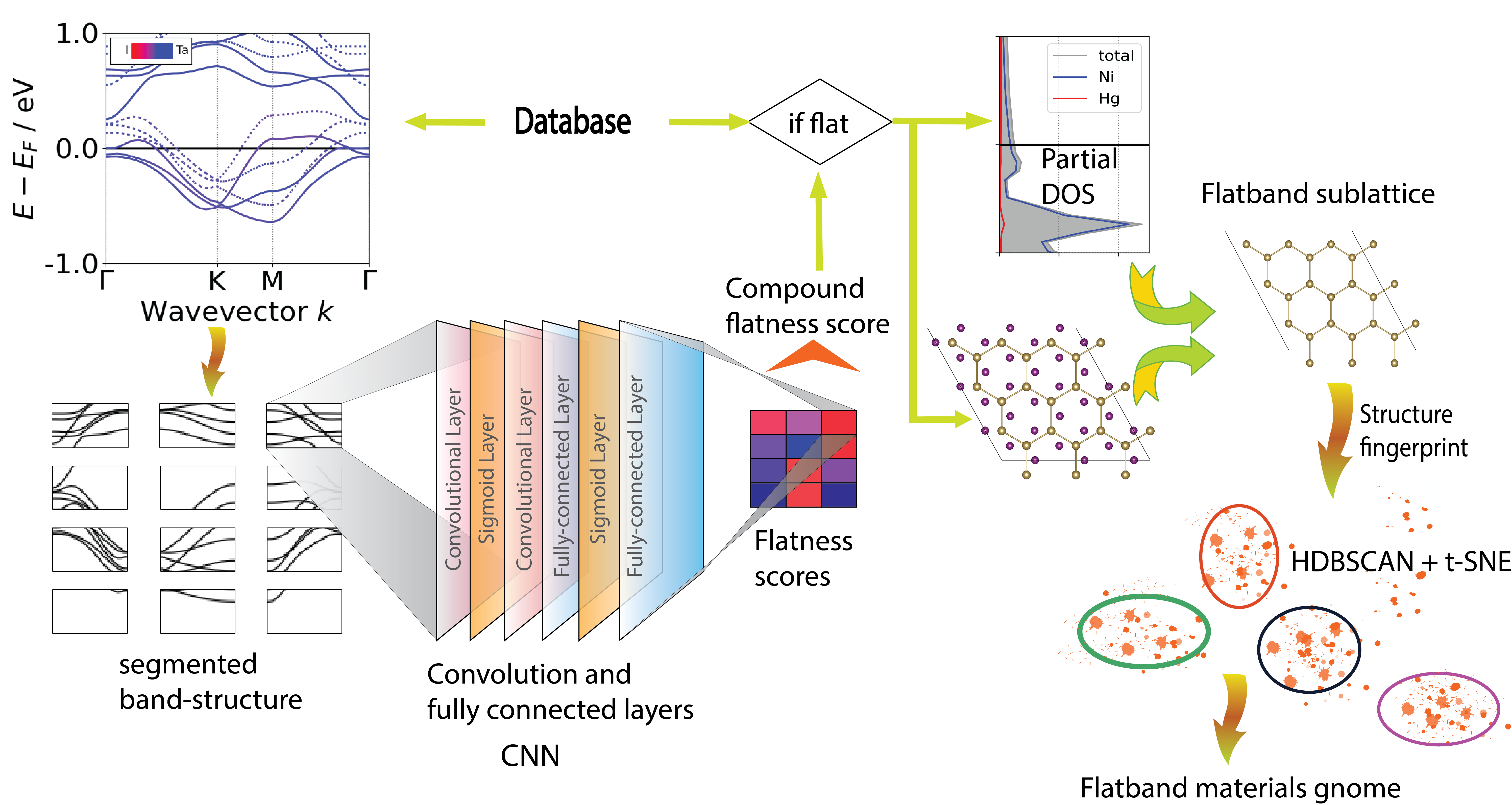}
    \caption{Architecture of combined supervised and unsupervised machine learning algorithms used in this work. CNN was trained to identify flat band materials using segmented band structure images from the database, followed by identifying sublattices that are responsible for flat dispersion using element projected DOS. Then, density-based clustering combined with t-SNE was used to classify the assigned structural fingerprints, and to identify  classes of flat band 2D materials.}
    \label{fig1}
\end{figure*}

Our protocol for the hybrid approach (supervised deep learning for flat band identification + unsupervised machine learning for symmetry-based clustering) is outlined in Fig. \ref{fig1}. First, CNN is trained and used to identify flat bands by examining all the 
band structure images across 2Dmatpedia, allowing us to reveal materials with flat bands spanning the whole Brillouin zone (BZ). Second, a crystal sublattice which has the maximum projected density of states (DOS) near the flat band is assigned as the sublattice responsible for the flat dispersion. Third, a structure descriptor algorithm CrystalNNFingerprint\cite{zimmermann2020local} is used to calculate structural similarity of the sublattices with predefined local coordination templates and assign a structure fingerprint. Finally, the vectorized fingerprints of the structures were fed into a bilayer unsupervised clustering module, with density-based hierarchical algorithm HDBSCAN\cite{campello2013density} as the inner layer and t-distributed Stochastic Neighbourhood Embedding\cite{vandermaaten08a} (t-SNE) as the output layer. This module clustered the flat band material population into stratified isostructural groups. All the developed code is available at \url{https://github.com/Anupam-Bh/ML_2D_flat_band}.

\subsection*{Identification of flat bands using CNN}
\begin{figure*}
    \centering
    \includegraphics[width=\textwidth]{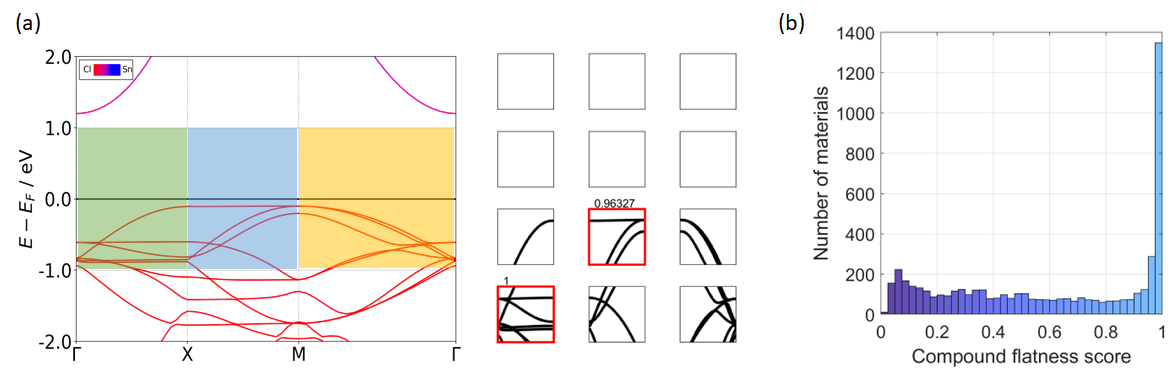}
    \caption{Identification of flat bands from band structure images using supervised machine learning. 
    (a) Segmentation of band structure images from 2Dmatpedia, and identification of flat band segments. Left panel: segmentation of band structure image of \ch{SnCl4} [2dm-13] horizontally into four 0.5 eV energy strips, and vertically along high symmetry points in k-space, as denoted in green, blue, and yellow for $\Gamma \rightarrow \mathrm{X}$, $\mathrm{X} \rightarrow \mathrm{M}$, and $\mathrm{M} \rightarrow \Gamma$ paths, respectively. Right panel: segmented band structure in the [-1,1] eV range and predicted outputs. Segments with flat bands are marked in red frames their corresponding flatness score is shown on top. (c) Histogram of compound flatness scores of all the materials in 2Dmatpedia.}
    \label{fig2}
\end{figure*}

For flat band recognition, we used CNN, which was trained on a subset of the Materials Project database, see Methods (Convolutional neural network). Figure \ref{fig2}a highlights one of the essential steps, image segmentation, which provides two main benefits. First, vertical segmentation allows identifying the location of a flat band with respect to the Fermi energy. Secondly, horizontal segmentation of the band structures at high-symmetry points enables us to find whether the dispersion is plane-flat (flat in the whole plane) or line-flat (flat only along one direction) across the BZ.

\begin{figure*}
    \centering
    \includegraphics[width=\textwidth]{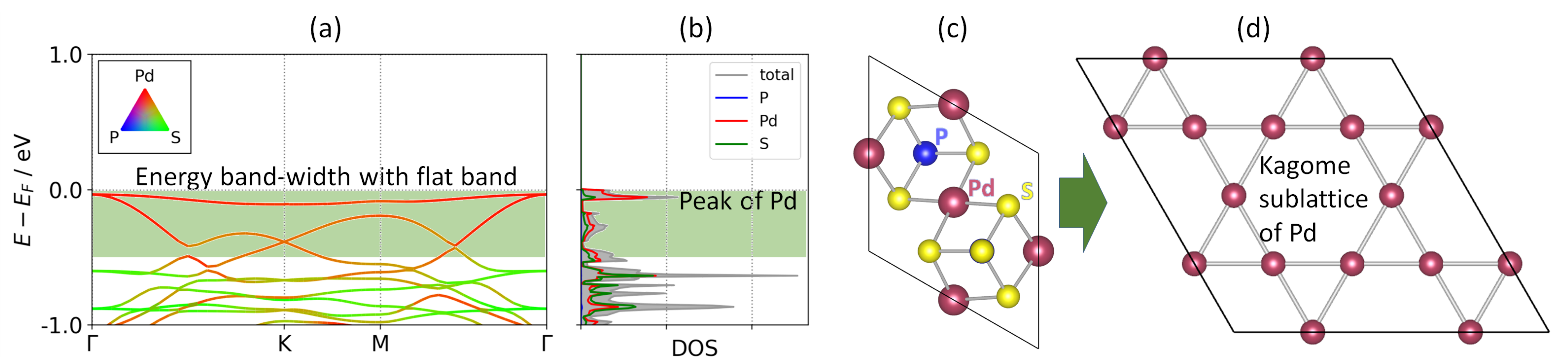}
    \caption{Identification of flat band sublattices. (a) Band structure of an identified flat band material example \ch{Pd3P2S8}. (b) Element projected DOS from (a) reveals the element responsible for flat band is Pd. (c) The crystal structure of \ch{Pd3P2S8}. (d) Kagome sublattice of Pd remains after stripping off P and S elements from \ch{Pd3P2S8} structure.}
    \label{fig3}
\end{figure*}

In this study we focused on plane-flat 2D materials, as their electrons have suppressed kinetic energy in wider momentum space, promoting electron-electron interactions. Figure \ref{fig3}a shows an example of plane-flat band material. For comparison, we show an example of line-flat band in Fig. \ref{fig2}a, where two flat segments (outlined in red boxes) belong to different energy strips. After all the materials with plane-flat bands have been identified (scored more than 0.5 along all the k-paths within an energy bandwidth), each material is assigned a compound flatness score between 0 to 1, which can be used to estimate the relative flatness of bands.

The number of materials as a function of their compound flatness score is shown in Fig. \ref{fig2}b, with a major peak near flatness score of 1. 
We found 2127 plane-flat materials out of total 5270 materials in 2Dmatpedia. 

\subsection*{Identification of flat band sublattices}
To build a connection between the identified flat band materials and their structural features, we propose a conjecture based on spatial localization of electrons forming a flat band. Let us assume a spatially localized region outside which electronic density approaches zero, formed either by the interference of wavefunctions or through other mechanisms. If this localized state originates from orbitals from multiple types of atoms, to keep the eigenvalue constant throughout the span of the localized state (plane-flat materials), there has to exist an accidental degeneracy between the states of those multiple types of atoms. Therefore, for most of the observed flat bands, the constituent electrons originate from a single element in the compound, assuming that the probability of accidental band degeneracy is low. Our conjecture is supported by several studies showing that it is the elemental sublattices which obey specific lattice and orbital symmetries that lead to flat bands \cite{kang2020topological,zhang2019kagome,nakai2022perfect, duan2022inventory}. For example, in intermetallic \ch{CoSn}, the flat band is attributed to the kagome sublattice of the transition metal element \cite{kang2020topological}, and in \ch{HgF2}, it is the diamond-octagon sublattice of mercury \cite{PhysRevMaterials.5.084203}. 

This conjecture allows us to use flat band element sublattice instead of full crystal structure for further analysis, largely simplifying the process while maintaining high accuracy. To this end, we first extract the element sublattice with the highest orbital contribution to the flat band.  
Here, element projected DOS (Fig. \ref{fig3}b) is used to identify the element with maximum density contribution, ignoring  the orbital mixing for classification purposes. As shown in Fig. \ref{fig3}, using flat band material \ch{Pd3P2S8} as an example, after the energy segment containing flat band is identified (green segment in Fig. \ref{fig3}a-b), the corresponding bandwidth in the element projected DOS is analyzed to obtain the element which has maximum projected DOS, and its sublattice. In this case, it is element Pd, and its sublattice, as shown in Fig. \ref{fig3}d. The lattice structure of \ch{Pd3P2S8} (Fig. \ref{fig3}c) is stripped off of P and S atoms to segregate the Kagome sublattice of Pd (Fig. \ref{fig3}d). As we conjecture, the symmetry operations relevant to this elemental sublattice lead to the flat band, thus only the sublattice with the chosen element for each compound is kept for further analysis. 

The extracted flat band sublattices are then subjected to a structure descriptor, CrystalNNFingerprint \cite{zimmermann2020local} and represented as a 244-dimensional vector to facilitate further classification using unsupervised machine learning algorithms,  
see Methods (Sublattice extraction and vectorization).

\subsection*{Unsupervised machine learning: bilayer clustering}
The vectorized flat band sublattices were further classified based on their structural fingerprints. We use a complementary bilayer classification algorithm to achieve optimal clustering. Density-based algorithm HDBSCAN \cite{campello2013density} is used to obtain hierarchical information and to identify closely-packed clusters,  while t-SNE\cite{vandermaaten08a} was compensating for the tendency of HDBSCAN to overlook local neighborhood information of the clusters. Details are given in the Clustering module section in Methods.

\begin{figure}
    \centering
    \includegraphics[width=0.5\textwidth]{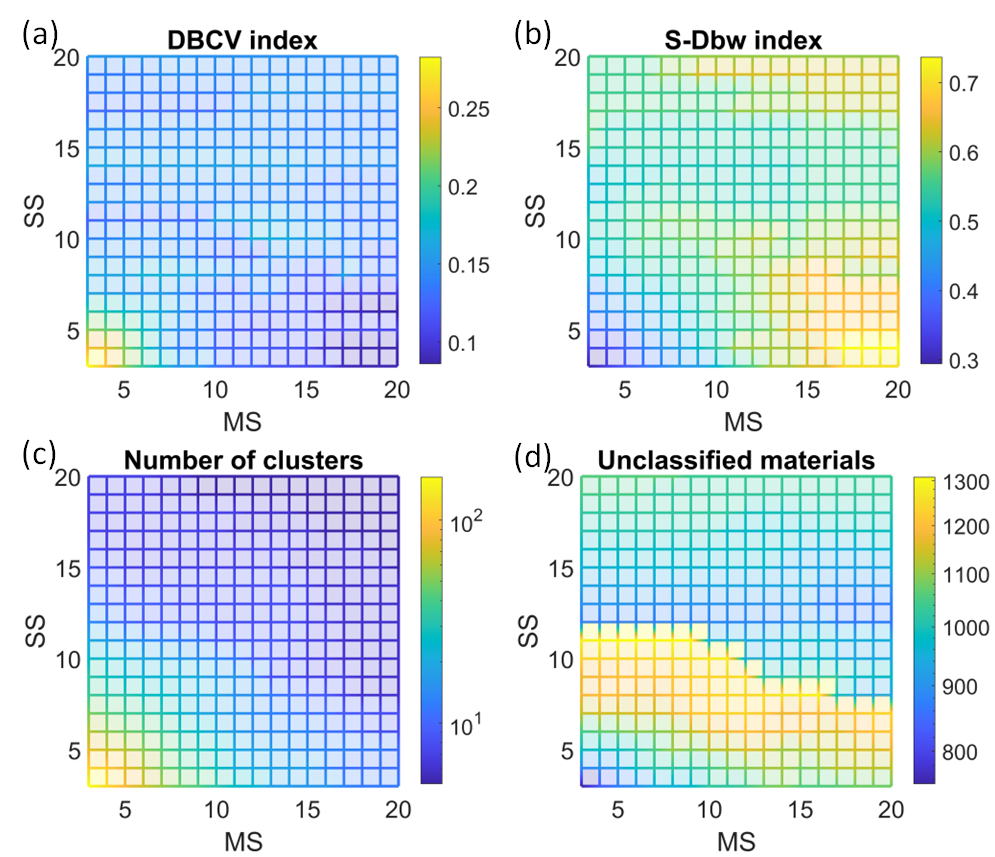}
    \caption{HDBSCAN algorithm optimization. Graphs showing (a) DBCV index, (b) S\_Dbw indices, (c) number of clusters and (d) number of unclassified materials as functions of minimum cluster size (MS) and sample size for density calculation (SS).}
    \label{fig4}
\end{figure}

Two parameters of HDBSCAN, minimum cluster size (MS), and sample size for density calculation (SS), were tuned to obtain optimal clustering solutions. The MS parameter trims off clusters which are smaller than its value and marks their members as unclassified, while the density around each point is calculated as a reciprocal relation to the distance of the point from its SS-th nearest neighbor. Both MS and SS were varied in a wide range, from 3 to 20, to obtain optimal clustering solutions. Performance of the clustering algorithm was quantified using two indices, density-based clustering validation (DBCV)  \cite{moulavi2014density},  and cluster  validity  index (S\_Dbw) \cite{989517}. DBCV calculates the intra-cluster and inter-cluster densities to estimate affinity between objects inside a cluster in comparison to connectivity between clusters. A higher DBCV index indicates a better clustering solution.  The S\_Dbw, on the other hand, is expressed as the sum of intra-cluster variance as a measure of cluster compactness, and inter-cluster density as a measure of separation. A smaller S\_Dbw index marks a better clustering solution.  Figure \ref{fig4}a-d shows the DBCV and S\_Dbw indices, the number of clusters, and the number of unclassified materials for different values of MS and SS. We notice that smaller values of MS and SS result in better DBCV and S\_Dbw, as well as a higher number of clusters. The number of unclassified materials in Fig. \ref{fig4}d, however, demonstrates a non-monotonic behavior with MS and SS: it first increases, caused by the existence of groups of fingerprints that are difficult to classify; then drastically drops, implying that many such dispersed fingerprints suddenly get included into  clusters, resulting in bad classification. Therefore, the choice of MS and SS parameters is expected to be smaller than the values at which such sudden change in Fig. \ref{fig4}d occurs, in order to obtain the optimal solution.

\begin{figure*}
    \centering
    \includegraphics[width=\textwidth]{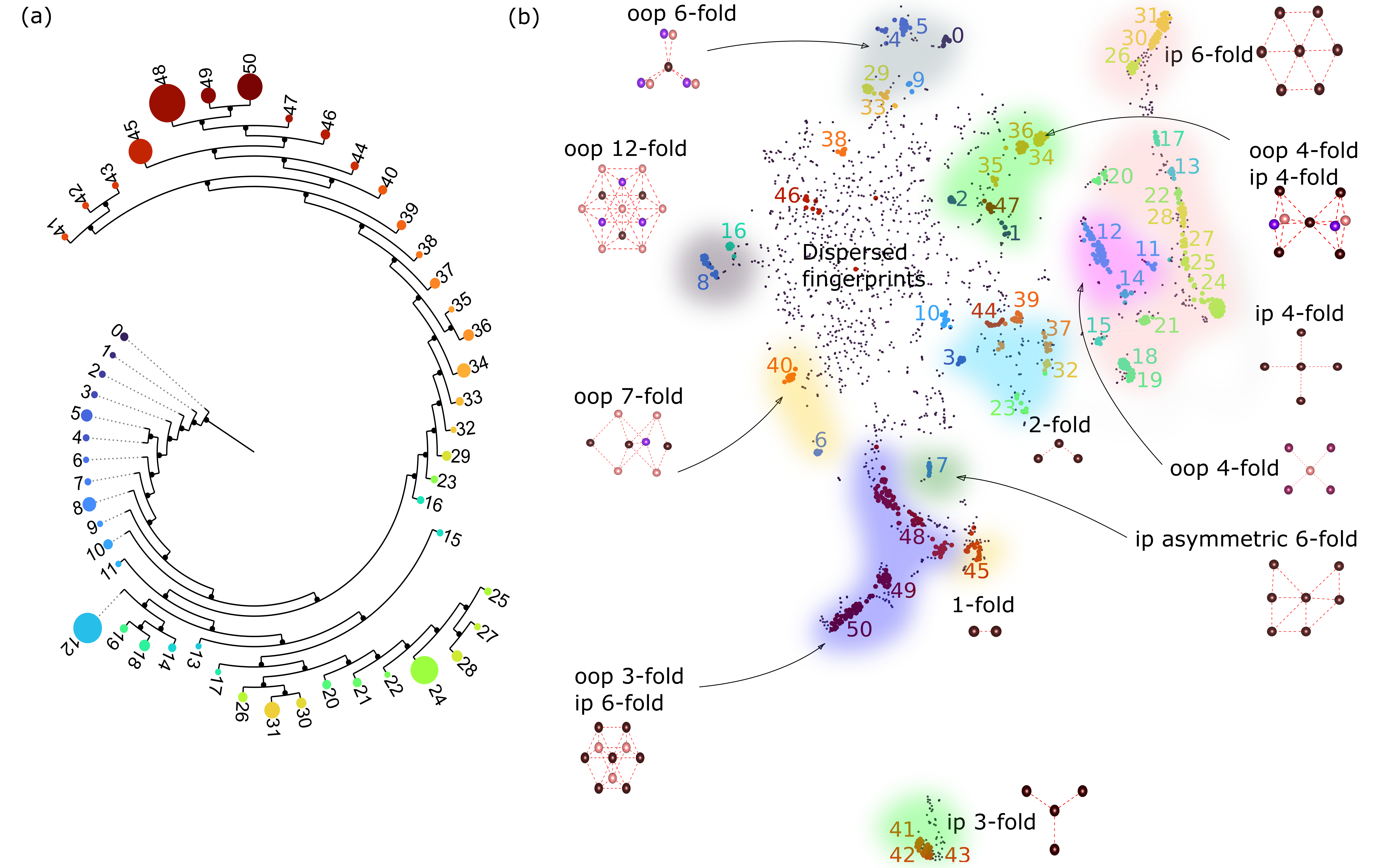}
    \caption{Clustering solution from HDBSCAN with MS = 7, SS = 6. (a) Phylogenetic tree expression of the hierarchical relations among clusters represented by colored circles. The size of each circle is proportional to the corresponding cluster's volume, while the cluster index is indicated by the number nearby. (b) t-SNE 2D visualization of the structure fingerprint space, different coordination patterns are color-coded; ip is in-plane and oop is out-of-plane coordination. Insets are representative coordination templates, where atoms in different atomic planes of the templates are colored differently.}
    \label{fig5}
\end{figure*}

For HDBSCAN, we choose MS = 7 and SS = 6 to map the coordination patterns into the structure fingerprint space, yielding a total of 51 clusters and 1448 unclassified materials. Our results are shown as a phylogenetic tree of clusters in Fig. \ref{fig5}a, and as a 2D t-SNE representation in Fig. \ref{fig5}b. A relatively small number of clusters facilitates the identification of major groups of structure fingerprints, although it yields a higher number of unclassified materials. We also tried to classify our results using a finer clustering solution with MS = 4 and SS = 3. Reduced MS and SS values improve the performance of the HDBSCAN and give a much higher number of clusters (131 clusters). Detailed classification of all the flat band sublattice structures for MS = 4 and SS = 3 is provided in Supplementary Information Fig. \ref{fig6} and Tables \ref{mono}-\ref{multilayer}.

Identifying structural groups from the HDBSCAN clusters has been largely possible because of the soft clustering feature of HDBSCAN. This density-based algorithm attaches a probability (representing the chance of being a member of a cluster; more about this membership probability in Methods) to each member of a cluster. The members with perfect probabilities or `exemplars' help the direct identification of sublattice represented by a cluster. Thus, we find exemplar sublattice structures from each cluster; however, often different clusters which are in close proximity, represent almost identical sublattices. To identify this local neighborhood, we use t-SNE. We combine the sublattice structures from nearly identical cluster exemplars to obtain the total list of lattice structures responsible for flat dispersion. For MS=7 and SS=6, we identify 28 such sublattice structures which are given in the Supplementary Note Table \ref{tab_7_6}. 

For the optimized t-SNE representation, distance information of the 90 nearest neighbors for each structure fingerprint is preserved using perplexity = 30. The principal component analysis is used to calculate the projection plane to retain maximal global neighborhood information. The learning rate is automatically adjusted according to  
\begin{equation}
    \mathrm{learning\ rate} = \frac{\mathrm{sample\ size}}{\mathrm{early\ exaggeration\ factor}}
\end{equation} 
where the early exaggeration factor is 12 for the first 250 iterations\cite{belkina2019automated, kobak2019art}. The total number of iterations was kept at 10000 to ensure convergence of t-SNE.

\subsection*{Evolution of the identified coordination patterns and lattice structures}

The phylogenetic tree in Fig. \ref{fig5}a shows the hierarchy among identified clusters of structurally similar lattices, which maps into the t-SNE plot in Fig. \ref{fig5}b.

By analyzing the clusters, 12 coordination patterns are revealed, shown as shaded regions in Fig. \ref{fig5}b. Fuzzy boundaries between different coordination patterns develop, as the coordination in each sublattice evolves gradually throughout the embedded space. Most well-defined isostructural clusters are found near the edges in the t-SNE plot, where the structure fingerprints provide a satisfactory description of underlying lattice geometries. On the contrary, many dispersed unclassified fingerprints appear closer to the central region in the t-SNE plot. This can be ascribed to lattices with large unit cells, which can be described by multiple coordination patterns (typically, more than 3--4) in their lattice-sites. The average of their multiply-coordinated structure fingerprints results in lattices that struggle to be properly included in any single coordination geometry. Examples of such dispersed fingerprints are seen in \ch{LiSb(PO3)4} with 24 oxygen sites, or \ch{AlPI8} with 16 iodine sites. Overall, finding an adequate structure descriptor for such large-unit-cell lattices is still an open problem\cite{himanen2020dscribe}. Some unclassified examples are also found lying near clusters despite sharing similar structures to those within the well-defined clusters, because of density-based segregation.

Let us highlight the main features of these plots. We first notice that clusters \#41-43, corresponding to honeycomb lattice materials with planar 3-fold coordination of atoms, clearly stand out at the end of the phylogenetic tree, Fig. \ref{fig5}a, and are positioned at the bottom of the t-SNE plot, Fig. \ref{fig5}b. Their separation from the rest of the charts marks the distinctness of their structural fingerprints. Moving to the top of Fig. \ref{fig5}b, we see the structure evolution from in-plane hexagonal sublattice (clusters \#26, \#30 and \#31) to out-of-plane hexagonal lattice (\#0) on the left, further to the out-of-plane 6-fold coordination sublattices (\#4, \#5, \#9 and \#29). Similarly, the transition from in-plane 4-fold coordination (e.g., \#22, \#27, and \#28) to out-of-plane 4-fold coordination (\#11, \#12, and \#14) occurs in the same manner. 

Our t-SNE chart also captures the gradual evolution between different coordination patterns, for example, from hexagonal to rectangular sublattices. Moving downwards from the in-plane hexagonal sublattice, we first encounter elongated hexagons (\#17), which can also be represented as planar centered orthorhombic lattice. Further distortion of the central atom creates planar monoclinic lattice (\#13 and \#20) and planar orthorhombic lattice (\#22, \#27, and \#28). From here, evolution into the square lattice (\#24 and \#25) is straightforward. Even the subtle structural differences, such as differences in stacking distance, within the same coordination are also well reflected in our clustering plot. For example, stacked hexagonal bilayer in 9-fold coordination (6-fold in-plane and 3-fold out-of-plane) with short stacking distance (\#45) is excluded from the branch for longer stacking distance coordination (\#48-50), as shown in the phylogenetic tree. Even among the clusters \#48-50, we could identify slight in-plane distortion in \#48 leading to coloring triangle lattices whereas \#49 and \#50 show no such distortion leading to simple AB stacked hexagonal lattice. Contrary to the well-defined isostructures, clusters that don't show any regular or identifiable coordination patterns (e.g., \#10, \#33, \#38, and \#46) lie around the dispersed region in the center of t-SNE plot.

\section*{Discussion}
Using our hybrid machine learning approach, we identified many sublattice structures responsible for hosting flat bands. Among these, some have already been confirmed in the literature, endorsing the validity of our approach. For example, kagome, breathing kagome, sawtooth, square-chain (Creutz) lattices are known to host flat bands \cite{kang2020topological,essafi2017flat,gremaud2017haldane, mondaini2018pairing}. But many of the lattices identified in this work are entirely not explored, and the origin of their flat bands remains to be clarified. Some of such monolayer sublattices are listed below, alongside example materials followed by their 2Dmatpedia id. These include Lieb-square lattice (\ch{Fe4S5} [2dm-4626], \ch{Os4S5} [2dm-1815]), centered orthorhombic lattice (\ch{Li4VF8} [2dm-4139], \ch{VZnF4} [2dm-4036]), diamond-chain lattice (\ch{ReS2} [2dm-4006], \ch{ReSe2} [2dm-3917]), $(3^2,4,3,4)$ Archimedean lattice (\ch{TiP} [2dm-2672]), vertex shared square chain lattice (\ch{Tl2Te5}[2dm-1601]), linear chain lattice (\ch{NbS3}[2dm-1482]), isolated quadrilaterals (\ch{MoSe2}[2dm-2007]) and centered monoclinic lattice (\ch{Ta2I5} [2dm-5407]). More examples are given in Table \ref{tab_7_6} and \ref{mono} in the Supplementary Information.

Bilayer flat band sublattices, in many cases, consist of stacked monolayer lattices with flat bands. For example, stacked kagome, stacked square, stacked centered orthorhombic, AB and AA stacked hexagonal, and stacked orthorhombic lattices belong to this category. Among bilayer sublattices, stacked centered-orthorhombic-square chains (\ch{SmF3} [2dm-875]) and isolated tetrahedra lattice (\ch{LiNiPO4} [2dm-5215], \ch{Si(HgO2)2} [2dm-4520]), are newly identified flat band lattices. 

Meanwhile, several few-layer sublattices made of stacked monolayer flat band sublattices were also identified, including stacked hexagonal and orthorhombic, kagome-hexagonal, and coloring-triangle-hexagonal lattices. Apart from this, we observed many interesting new multilayer sublattices with flat bands. For example, sublattices that consist of in-plane octahedras, where different octahedra connections generate seven new sublattices. These are vertex sharing octahedra chain (\ch{YFeF5} [2dm-3862]), edge sharing octahedra-parallel chain (\ch{Sc5Cl8} [2dm-4448]), vertex sharing planar octahedra (\ch{ZrF4} [2dm-5172]), vertex-edge sharing planar octahedra (\ch{CaTlCl3} [2dm-4337]), isolated twisted octahedra (\ch{ReSeCl} [2dm-5461]), edge-sharing zigzag octahedra chain (\ch{ZnMoO4}[2dm-3090]) and edge sharing octahedra chain (\ch{Y2Cl3} [2dm-3786]). Furthermore, the stacked $(3^2,4,3,4)$ Archimedean sublattice (\ch{CoBr4} [2dm-1906], \ch{MnF4} [2dm-5281])with out-of-plane distortion, is also identified to host flat bands. Other new multilayer sublattices, e.g. vertex  sharing triangular pyramid chain (\ch{V2CuO6} [2dm-4387]), triangular bipyramid lattice (\ch{MoO3} [2dm-5495]), distorted Kagome lattice (\ch{MgCl2} [2dm-4672]), planar hexahedra lattice (\ch{TaF5} [2dm-4011]), and stacked orthorhombic-square (\ch{HgI2} [2dm-3966]) 
are also found to show flat bands. Future research is expected to reveal the origin and the rich physics behind the majority of the potential flat band lattices listed here.

Our hybrid approach combining supervised and unsupervised machine learning has successfully demonstrated its accuracy and efficiency in navigating through enormous information space to hunt for new flat band materials. This approach can be conveniently adapted in searching for materials with intended properties, and to mitigate issues arising from the quality of high-throughput density functional theory calculations in existing databases. An interesting direction is to identify topologically nontrivial flat bands, which show crossings with other dispersive bands at high symmetry points in the BZ \cite{rhim2019classification}. Features like band crossings, which are both common and small, may require developing automatic graphical pattern search tools\cite{Borysov2018}. Moreover, in addition to using structural fingerprints, such as lattice coordination patterns as described in the current work, to distinguish different flat band materials, electronic fingerprints, e.g., the similarity of electronic states\cite{knosgaard2022representing}, may also be attempted in the future to offer new perspectives towards the discovery of new materials.

\section*{Methods}
\subsection*{Convolutional neural network}
The CNN trained for the identification of flat bands consists of 6 layers (as shown in Fig. \ref{fig1}) and 6.9 million learnable parameters. It is a regression network which predicts the existence of a flat band in the image segment and outputs as a single value -- the flatness score. The resolution of each segment is set to $96\times96$ pixels in consideration of both image quality and computing time. The first convolutional layer consists of 30 channels of $10\times10$ filters with [1,1] stride. The second convolutional layer has 12 channels of $3\times3$ filters with [1,1] stride. Sigmoid activation was used and found to give better performance than softmax and ReLU activation as the output of the network lies within [0,1]. The first fully-connected layer has 80 output nodes, whereas the second has a single output. For training purposes, Adam\cite{kingma2014adam} optimization algorithm was used with learning rate $5\times10^{-5}$ and L2-regularization parameter 0.003.

To train the CNN, band structures of approximately 1900 materials from the Materials Project database were used. There are several reasons for selecting the Materials Project database for training. First, it offers more training flexibility compared to the relatively small database of 2Dmatpedia ($\approx 5300$ materials), avoiding applying the trained algorithm back to the same database. Second, training the algorithm on 3D compounds that have inherently more complicated band structures could improve its accuracy when applied to the 2Dmatpedia. Finally, training the model on a general 3D database allows future application to many other available materials databases.

The developed CNN model was then tested on 7000 image segments from Materials Project. The test set predictions match 92.1$\%$ with manual selection, whereas false negative and false positive predictions were seen in 3.3$\%$ and 4.6$\%$ of all cases, respectively. This is quite high precision since even manual selection of flat segments could not guarantee 100$\%$ consistency, as partially flat segments can often lead to confused judgment from a human selector.

Before applying the trained CNN to 2Dmatpedia, band structures from 2Dmatpedia in the energy range [-1,1] eV were segmented at high symmetry points of k-space. Each resulting segment has $96\times96$ resolution. For example, the band in Fig. \ref{fig2}b is divided horizontally into four 0.5 eV energy bandwidths, and vertically along k-path into three segments (as denoted in green, blue and yellow). A threshold flatness score predicted from the trained CNN is used to determine whether a band-structure segment is flat (score $\geq  0.5$) or non-flat (score $< 0.5$). We find a significantly higher fraction of flat band segments in 2Dmatpedia, approximately 25$\%$, than that in 3D materials from Materials Project (approximately 13\%). Dimensionality reduction and accompanied restriction in degrees of freedom of electrons in 2D materials are likely to be the reason. To identify plane-flat band
2D materials, which have flat bands throughout high symmetry lines and are of particular interest, we select the energy strip in which all the horizontal segments are recognized as flat. In the rare cases where multiple energy strips contain all-flat segments, priority is given to the energy bandwidth where electrons are more easily accessible, i.e. closer to the Fermi level. 

\subsection*{Sublattice extraction and vectorization}
We use the element projected DOS, as shown in Fig. \ref{fig3}a-b, to identify the element corresponding to the flat band sublattice. Because flat electronic bands are accompanied by corresponding peaks in DOS, an element with maximum density contribution in the flat band is most likely to be responsible for the flat dispersion, although orbital mixing among different species is also a contributing factor. Here we conjecture that the symmetry operations pertaining to this elemental sublattice lead to the flat band, thus only the sublattice with the chosen element for each compound is kept for further analysis. After the energy segment containing flat band is identified, the corresponding bandwidth in the element projected DOS is analyzed to obtain the element and, subsequently, its sublattice. 

A structure descriptor was then applied to a selected sublattice to create a vector representation for machine learning algorithms. Here we use a structure descriptor CrystalNNFingerprint \cite{zimmermann2020local}, implemented in the Matminer package \cite{ward2018matminer}. Other structure descriptors, like Coulomb matrix, Ewald sum matrix, Smooth Overlap of Atomic Positions (SOAP), Many-body Tensor Representation (MBTR), usually yield a constant length vector representing the structure \cite{himanen2020dscribe,bartok2013representing,lee2021descriptors}, were found less suitable for our analysis. CrystalNNFingerprint retains more information when the number of atomic sites is high, thus particularly suitable for identifying local coordination patterns in crystals even in the presence of small lattice distortions. For example, popular descriptors like SOAP calculating spectral average of the local fingerprints, would fail to retain local structural information. CrystalNNFingerprint first determines the local environment for each atomic site using a neighbor-finding algorithm based on Voronoi decomposition. The resulting coordination pattern is then compared with Local Structure Order Parameters (LoStOPs) which are different coordination templates, and a 61-dimensional fingerprint vector is assigned to the atomic site. The global fingerprint for the whole structure is then represented by arranging the mean, standard deviation, maximum and minimum of each local (atomic site) CrystalNNFingerprint within a structure. The final structure fingerprint lies in a 244-dimensional vector space. 

\subsection*{Clustering module}
We then used unsupervised machine learning module consisting of density-based clustering algorithm HDBSCAN \cite{campello2013density} and t-SNE \cite{vandermaaten08a} to further cluster 244-dimensional structure fingerprints. HDBSCAN creates clusters with fingerprints that are densely populated, acting as a strict identifier of similarity in coordination patterns with many fingerprints designated as outliers. Simultaneously, soft clustering feature of HDBSCAN yields a probability for each material to be included in a cluster. Consequently, members of clusters with very high inclusion probability (exemplars) facilitate straightforward identification of corresponding sublattice structures. The exemplars are the members of a cluster which persist in the cluster for the largest range of density. Hence, these may be identified as the central members even if the shape of the cluster is fairly complicated as they do not depend on any distance metric. However, HDBSCAN tends to assign fingerprints with  even small discrepancies into different clusters, risking overlooking local neighborhood information. As a complement, we use t-SNE to visualize and identify similarities among HDBSCAN clusters, acting as a second layer of unsupervised classification. The nonlinear dimensionality reduction capability of t-SNE allows both local neighborhood information and global distance relations to be preserved, in comparison to other algorithms like Isomap \cite{tenenbaum2000global}, Locally Linear Embedding (LLE) \cite{roweis2000nonlinear}, Hessian eigenmapping \cite{donoho2003hessian}, etc. We also tested UMAP \cite{mcinnes2018umap}, no significant improvement over t-SNE is observed (results are discussed in the Supplementary Information section II). 

The choice of HDBSCAN as the first layer of unsupervised clustering algorithm is based on the fact that more commonly utilized clustering algorithms, like K-means and hierarchical clustering, would yield non-optimum segregation. K-means algorithm solely depends on the distance among inputs, generating good results only for inputs that form ellipsoid clusters. At the same time, hierarchical clustering is too sensitive to the presence of noise in the data, making it unsuitable in this context since lattice distortion is an avoidable source of noise. For our 244-dimensional data we used Manhattan distance metric, $L_{1}$ norm, since Euclidean distance metric ($L_2$ norm) becomes less effective in calculating distances among points in high-dimensional spaces \cite{aggarwal2001surprising}.

\section*{Data Availability}
The datasets generated during and/or analyzed during the current study are available from the corresponding author on reasonable request.
\section*{Code Availability}
The codes generated during the current study are available in the Github repository, \texttt{https://github.com/Anupam-Bh/ML\_2D\_flat\_band}.
\section*{Author contributions}
A.B. carried out the calculations, implemented the algorithms in codes, analyzed the data and drafted the manuscript. A.M. conceived the research plan, analyzed the data, drafted the manuscript and supervised the research work. R.C. co-supervised the research work. I.T. and Q.Y. analyzed the data and contributed to writing the manuscript.
\section*{Acknowledgements}
This research was supported by the European Research Council (ERC) under the European Union's Horizon 2020 research and innovation program (Grant Agreement No. 865590), and the Royal Society International Exchanges 2019 Cost Share Program (IEC\textbackslash R2\textbackslash 192001). A.B. acknowledges the Commonwealth Scholarship Commission in the UK for financial assistance. Q.Y. acknowledges the funding from Leverhulme Early Career Fellowship ECF-2019-612, Dame Kathleen Ollerenshaw Fellowship from the University of Manchester, and Royal Society University Research Fellowship URF\textbackslash R1\textbackslash 221096. 
\section*{Competing Interests}
The authors declare no Competing Financial or Non-Financial Interests.

\clearpage

\renewcommand{\thepage}{S\arabic{page}} 
\renewcommand{\thesection}{S\arabic{section}}  
\renewcommand{\thetable}{S\arabic{table}}  
\renewcommand{\thefigure}{S\arabic{figure}}
\setcounter{figure}{0}
\setcounter{section}{0}

\section*{Supplementary Information}

\begin{longtable*}{|p{0.15\textwidth}|p{0.36\textwidth}|m{0.2\textwidth}|p{0.24\textwidth}|}
    \caption{Flatband sublattices obtained from Fig.\ref{fig5}} 
    \\\hline  
    Lattice type & Cluster ID\#[cluster size]:Example[sublattice] & Remarks & Structure \\\hline
    1.Monolayer hexagonal lattice & \makecell{0[10]: \ch{ Rb2O [O]}, \ch{Ti2S [S]}\\26[12]: \ch{CrPb3 [Cr]}, \ch{Hg3AsS4Br [As]} \\ 30[13]: \ch{Co(ReO4)2 [Co]}, \ch{BaCr4O8 [Ba]}\\ 31[21]: \ch{CdClO [O]}, \ch{MnCl2 [Mn]}} &    &\includegraphics[height=0.6in]{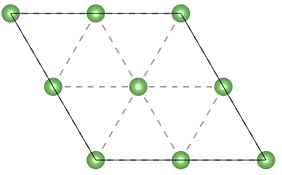} \\\hline
    2.Sawtooth-chain lattice  &  \makecell{32[7]: \ch{Ta2TiZn2O8 [Ti]}, \ch{Ta2TeO8[Te]} \\ 3[8]:\ch{V2ZnS5[V]}, \ch{V2ZnO5[V]} \\ 23[9]:\ch{CuS4BrN4 [Cu]} \\ 44[10]: \ch{BaI2 [I]} \\ 45[33]: \ch{PdI2 [Pd]}, \ch{NiI2 [Ni]} } &    & \includegraphics[height=0.6in]{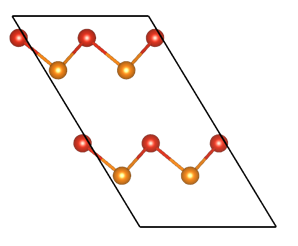} \\\hline
    3.Planar Centered monoclinic lattice   &  \makecell{15[8]: \ch{K2Hg3(GeS4)2 [K]}, \ch{Ta2I5 [Ta]}, \ch{VSeO4[V]}} &    & \parbox[c]{1em}{\includegraphics[height=0.6in]{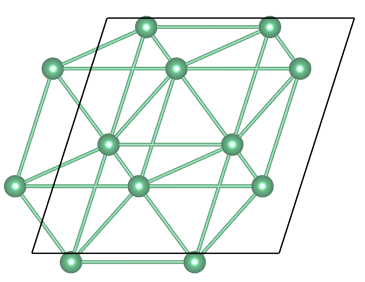}} \\\hline
    4.Planar centered-orthorhombic/ monoclinic lattice & \makecell{13[7]: \ch{Co(SiSe2)2 [Co]}, \ch{Zn(SbO2)4 [Zn]}\\ 17[7]: \ch{Cu2IO3 [Cu]}\\ 20[11]:\ch{Li4VF8 [V]}, \ch{Tl2Pt(CN)4 [Pt]}\\  21[11]: \ch{ReF4 [Re]}, \ch{VZnF4 [V]} }& Each atom has planar 6-fold coordination (elongated hexagonal) &  \parbox[c]{1em}{
      \includegraphics[height=0.6in]{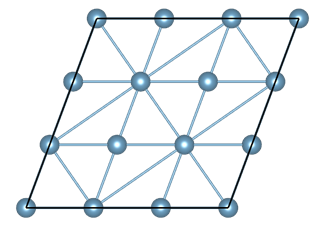}} \\\hline
    5.Kagome and breathing-Kagome lattice &  \makecell{19[10]: \ch{P2Pd3S8 [Pd]}, \ch{FeCo3O8 [Co]}\\ 18[14]:\ch{Nb3SBr7 [Nb]}, \ch{Nb3TeI7 [Nb]}} &  19: Kagome lattice 18: Breathing Kagome lattice & \parbox[c]{1em}{\includegraphics[height=0.6in]{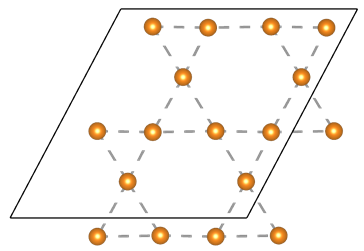}} \\\hline
    6.Planar orthorhombic lattice  &  \makecell{22[7]: \ch{LiVF4 [V]}, \ch{LiVCuO4 [Cu]}\\ 25[9]: \ch{LiAuF4 [Au]}, \ch{MnBiSe2I [Mn]}\\27[8]: \ch{CrPS4 [Cr]}, \ch{Cu(IO3)2 [Cu]}\\ 28[14]: \ch{Li3Co(NiO3)2 [Co]}, \ch{AlNiF5 [Al]}}  &   & \parbox[c]{1em}{\includegraphics[height=0.6in]{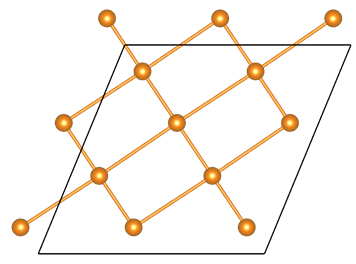}} \\\hline
    7.Monolayer square lattice  &  \makecell{24[38]: \ch{CuBr [Cu]}, \ch{LaNb2O7 [La]}}  &    & \parbox[c]{1em}{\includegraphics[height=0.6in]{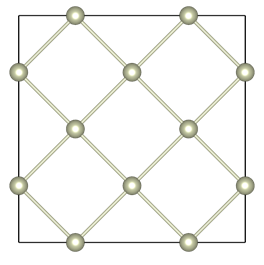}} \\\hline
    8.Honeycomb lattice & \makecell{41[8]: \ch{CrGeTe3 [Cr]}, \ch{TiI3 [Ti]}\\ 42[8]:\ch{IrBr3 [Ir]}, \ch{CrBr3 [Cr]} \\ 43[8]:\ch{RhCl3 [Rh]},\ch{VCl3 [V]} }&  &  \parbox[c]{1em}{
      \includegraphics[height=0.6in]{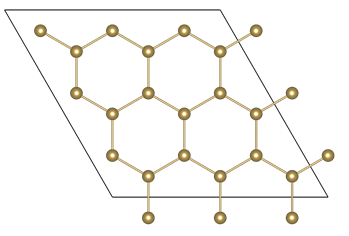}} \\\hline
    9.Lieb-square lattice  &  \makecell{7[8]: \ch{Fe4S5}, \ch{Os4S5}}  &  Lattice consists of Lieb-square pattern; each atom is asymmetric 6-fold coordinated  & \parbox[c]{1em}{\includegraphics[height=0.6in]{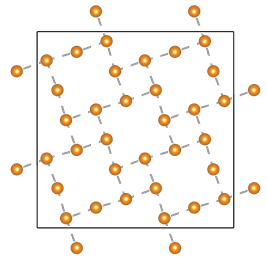}} \\\hline
    
    10.Diamond-chain lattice   &  \makecell{9[7]:\ch{ReSe2[Re]}, \ch{ReS2[Re]}}  &    & \parbox[c]{1em}{\includegraphics[height=0.6in]{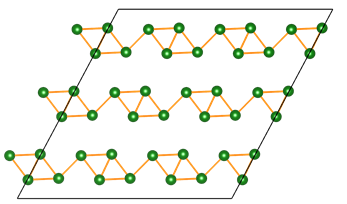}} \\\hline
    11. Isolated tetrahedra lattice  &  \makecell{37[13]: \ch{LiNiPO4 [O]}, \ch{Si(HgO2)2 [O]}, \ch{Cd2SiO4 [O]} }  &  Each atom has 3 fold coordination   & \parbox[c]{1em}{\includegraphics[height=0.8in]{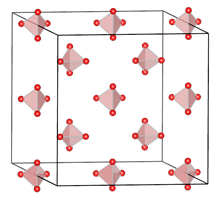}} \\\hline
    12.Face sharing cuboctahedron & \makecell{1[7]: \ch{Sr4Mn3(ClO4)2[O]}, \ch{Ba2Cr3O8 [O]}} & & \parbox[c]{1em}{\includegraphics[height=0.6in]{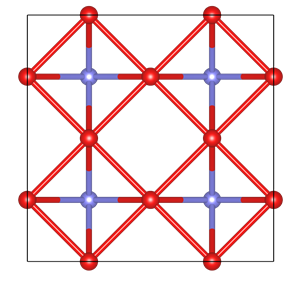}} \\\hline
    13.Vertex sharing octahedra chain  &  \makecell{2[8]: \ch{YFeF5[F]}, \ch{YWF5[F]}}  &    & \parbox[c]{1em}{\includegraphics[height=0.6in]{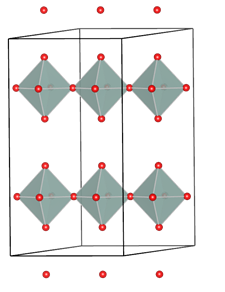}} \\\hline
    14.Edge sharing octahedra-parallel chain  &  \makecell{4[8]:\ch{Sc5Cl8[Sc]},	\ch{Sc7Cl10[Sc]}}   &   & \parbox[c]{1em}{\includegraphics[height=0.6in]{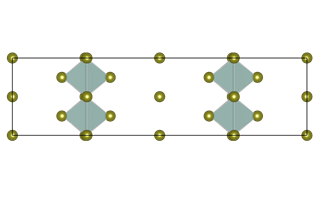}} \\\hline
    15.Distorted Kagome lattice  &  \makecell{5[15]:\ch{MgCl2[Cl]}, \ch{MgF2[F]}}  & Hexagonal lattice stacked over Kagome to make it distorted   & \parbox[c]{1em}{\includegraphics[height=0.6in]{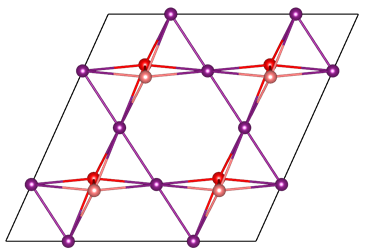}} \\\hline
    16.Stacked Kagome-hexagonal lattice  &  \makecell{6[7]:\ch{Zr(ReO4)2[O]}, \ch{Mg(AlH4)2[H]}}  &  Two Kagome atomic layers stacked with 2 hexagonal lattice layers at two ends  & \parbox[c]{1em}{\includegraphics[height=0.6in]{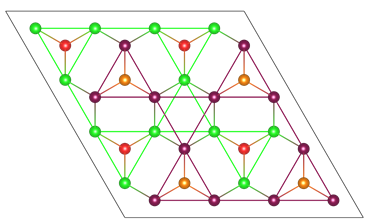}} \\\hline
    17.Dice lattice  &  \makecell{8[18]:  \ch{HgSe [Se]} }  &  Layers with ABCA.. stacking : atoms from inside layers have 12 fold coordination: form Triangular gyrobicupola  ; surface atoms have 9 fold coordination  & \parbox[c]{1em}{\includegraphics[height=0.6in]{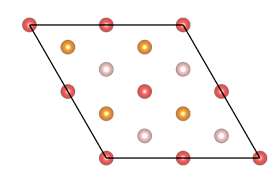}} \\\hline
    18.Stacked square (AB stacking) lattice  &  \makecell{ 11[7]: \ch{InClO[Cl]}, \ch{CaClF[Cl]} \\ 12[40]: \ch{LiFeF4[F]}, \ch{MoPO5[Mo]}\\ 14[10]:\ch{BaMgsn [Sn]}, \ch{BaMgPb [Pb]} }  &  Two square lattices stacked with AB stacking & \parbox[c]{1em}{\includegraphics[height=0.6in]{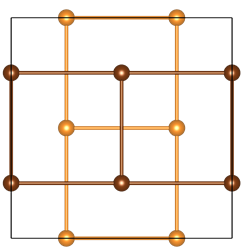}} \\\hline
    19.Triangular bipyramid  &  \makecell{29[12]: \ch{MoO3[Mo]}, \ch{SbBr3[Br]}}  &  Three layers with middle layer atoms having 6 fold coordination and side atoms having 3 fold coordination.  & \parbox[c]{1em}{\includegraphics[height=0.6in]{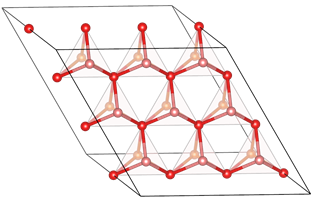}} \\\hline
    20.Planar hexahedra  &  \makecell{33[10]:\ch{TaF5[F]} \ch{SbF5[F]}}  &    & \parbox[c]{1em}{\includegraphics[height=0.6in]{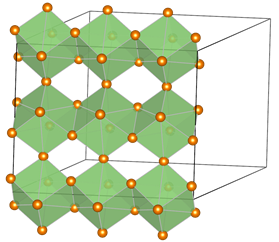}} \\\hline
    21. Vertex sharing planar octahedra  &  \makecell{34[18]: \ch{ZrF4[F]}, \ch{SnF4[F]} }  &    & \parbox[c]{1em}{\includegraphics[height=0.6in]{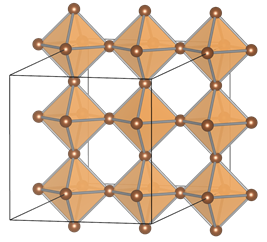}} \\\hline
    22.Stacked hexagonal (AB stacking)  &  \makecell{49[20]: \ch{Mn3CrO8[O]}, \ch{Mn3CuO8[O]} \\  50[35]: \ch{LaBr[La]}, \ch{YCl [Y]}}  &    & \parbox[c]{1em}{\includegraphics[height=0.6in]{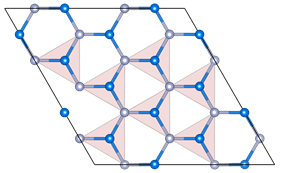}} \\\hline
    23.Stacked coloring triangle lattice  &  \makecell{ 48[51]:  \ch{InBr3[Br]}, \ch{GaBr3[Br]} }  &   & \parbox[c]{1em}{\includegraphics[height=0.6in]{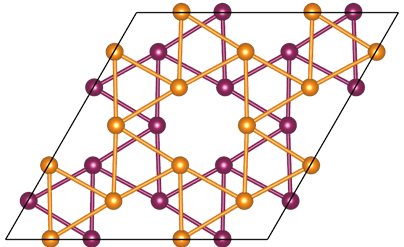}} \\\hline
    24.Stacked breathing Kagome  &  \makecell{39[12]: \ch{InAs3 [As]} }  &  Can also be called isolated octahedra   & \parbox[c]{1em}{\includegraphics[height=0.6in]{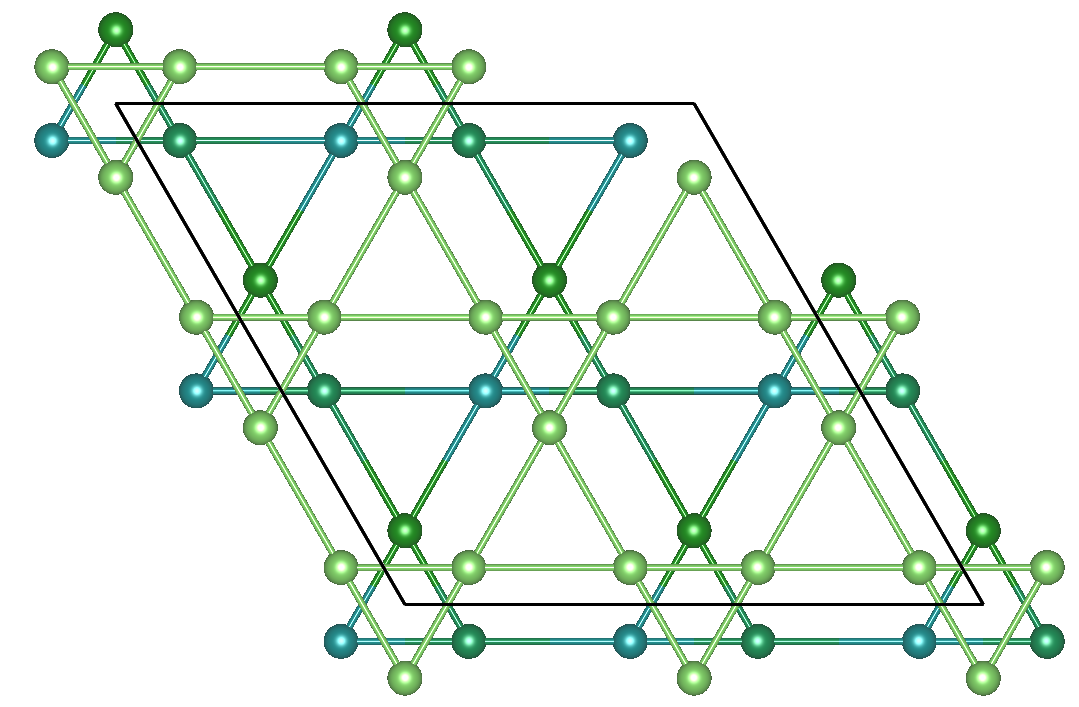}} \\\hline
    25.Stacked breathing coloring triangle  &  \makecell{35[7]: \ch{Ge3Sb2O9 [O]} \ch{Si3Bi2O9 [O]} }  &     & \parbox[c]{1em}{\includegraphics[height=0.6in]{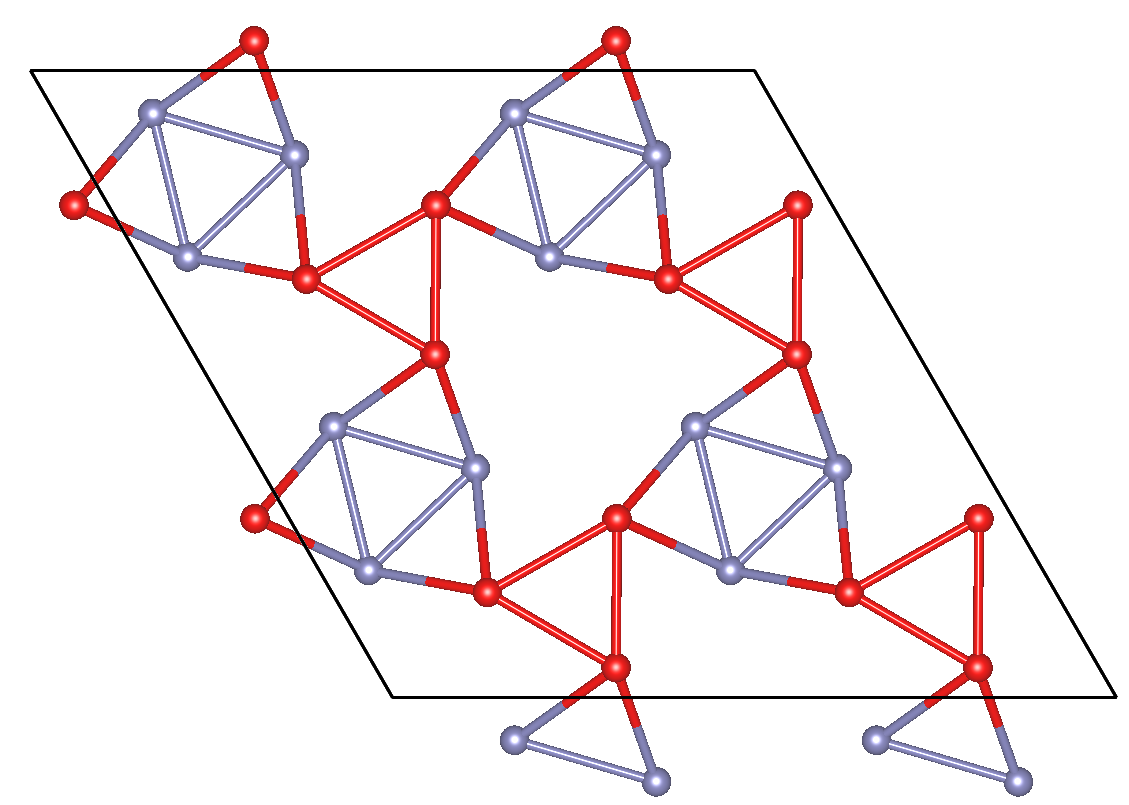}} \\\hline
    26.Stacked orthorhombic-square lattice  &  \makecell{36[14]:\ch{HgI2[I]}}  &  The cationic sublattice is a diamond octagon sublattice: But the flat-bands come from the anionic sublattice (stacked orthorhombic-square lattice)  & \parbox[c]{1em}{\includegraphics[height=0.6in]{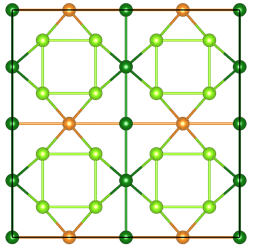}} \\\hline
    27.Edge-sharing octahedra chain  &  \makecell{40[11]: \ch{Y2Cl3[Y]}}  &    & \parbox[c]{1em}{\includegraphics[height=0.8in]{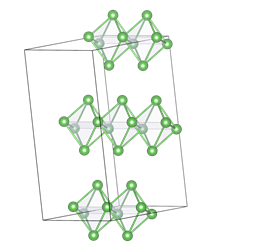}} \\\hline
    
    \label{tab_7_6}
\end{longtable*}

\subsection*{Supplementary Note I: Finer search for recurrent lattice structures}
\begin{figure*}
    \centering
    \includegraphics[width=\textwidth]{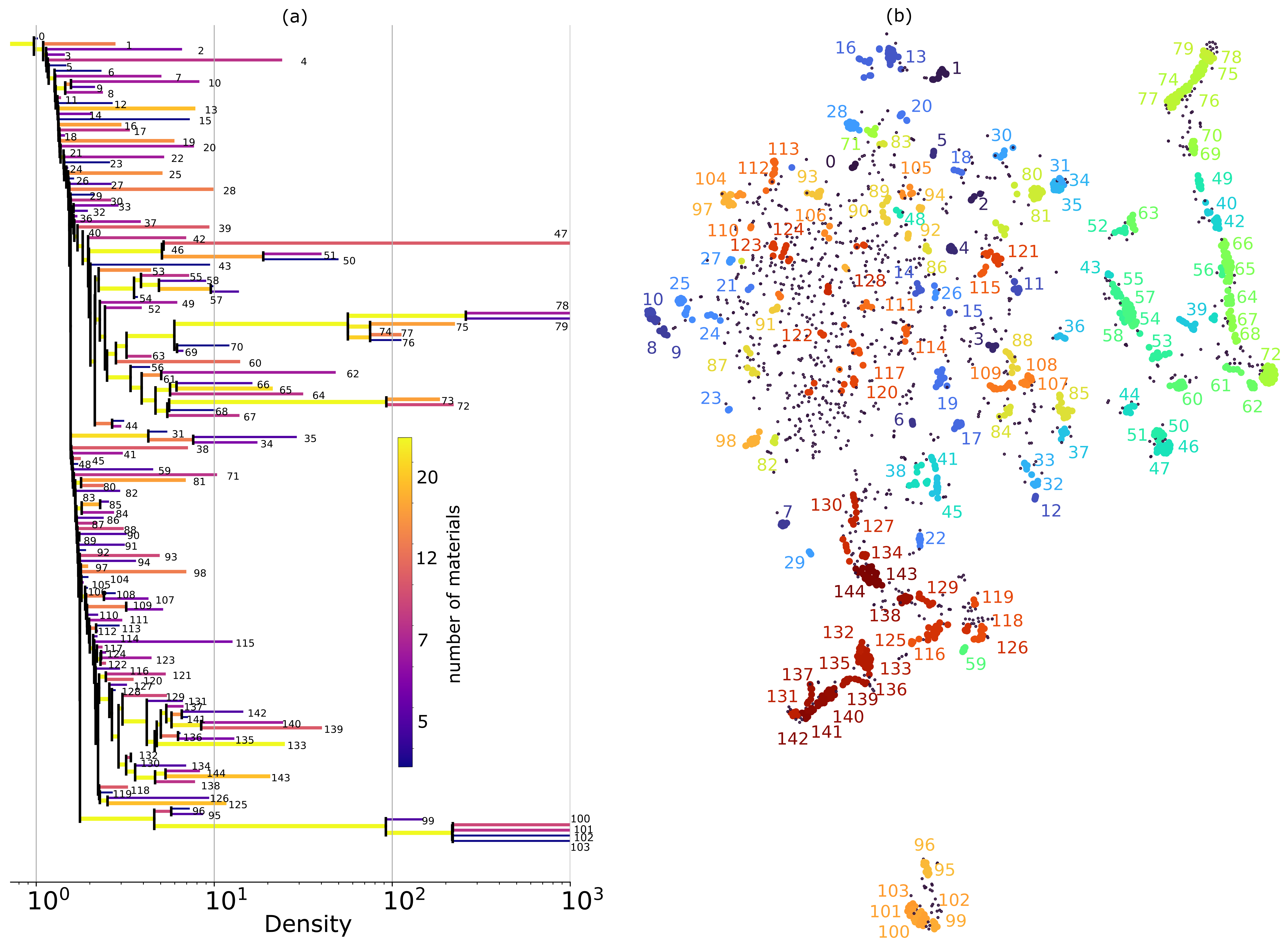}
    \caption{Clustering results from HDBSCAN for MS=4, SS=3. (a) Condensed tree plot showing hierarchical information among clusters with numbers representing cluster IDs. (b) t-SNE plot showing adjacency of materials in the embedded space. Numbers represent cluster IDs.}
    \label{fig6}
\end{figure*}

We also use a finer clustering solution by HDBSCAN which identifies more flat band sublattices as compared to those observed with MS = 7 and SS = 6 in Fig. \ref{fig5} and Table \ref{tab_7_6}. As demonstrated in Fig. \ref{fig4}a-b, the clustering solution for MS = SS = 3 yields the best DBCV and S-Dbw indices. However, this also means a much larger number of clusters (230), inducing extra complexity in identifying major groups of structure fingerprints. Instead, we choose MS = 4, SS = 3 to obtain a total of 145 clusters while maintaining a good number of unclassified materials (1071), comparing to other combinations like MS = 3, SS = 4 or  MS = SS = 4, which lead to a larger number of uncategorized examples. Comparing to the case of MS = 7, SS = 6, we now have smaller clusters, many of which were identified as outliers before. Results of HDBSCAN clustering with MS = 4 and SS = 3 are shown as a condensed tree plot in Fig. \ref{fig6}a and as a 2D t-SNE visualization in Fig. \ref{fig6}b. All the recognizable groups have been identified as clusters, and the number of unclassified examples around defined clusters is reduced.

We identified all the exemplar lattice structures found in the 145 HDBSCAN clusters shown in Fig. \ref{fig6}. The coordination patterns from the t-SNE representation in Fig. \ref{fig5}b could guide the identification/nomenclature of these structures. We again use t-SNE to combine the nearly identical exemplar candidates and obtain the total list of sublattice structures. Most sublattice structures of this list were identified with MS=7, SS=6 from Fig. \ref{fig5} and are already listed in Table \ref{tab_7_6}. New structures which were elusive with MS=7, SS=6 are described in detail in Tables \ref{mono}-\ref{multilayer}. We observe that most of the clusters are homogeneous as a common sublattice is present in all the materials within one cluster. However, 24 clusters were found to have either inhomogeneous sublattice structures or structures which are too complex to be categorized;  these clusters lie near the diffused region in the middle of t-SNE plot, Fig. \ref{fig6}b.

\begin{longtable*} {|p{0.15\textwidth}|p{0.36\textwidth}|m{0.2\textwidth}|p{0.24\textwidth}|}
      \caption{New monolayer flatband sublattices identified from Fig \ref{fig6}} 
      \\ \hline  
      Lattice type & Cluster ID\# [cluster size]:Example [sublattice] & Remarks & Figure \\
      \hline   
      1.Square chain lattice (Cruetz lattice)  &  \makecell{12[4]:\ch{KFe2Se3[Fe]}, \ch{RbFe2Te3 [Fe]}}  &  Spaced rows of squares  & \parbox[c]{1em}{\includegraphics[height=0.8in]{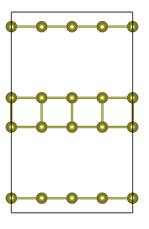}} \\\hline
      2.Vertex shared square chain lattice  &  \makecell{3[6]:\ch{Be2Te7Cl6[Te]}, \ch{Tl2Te5 [Te]}}  &  Spaced rows of squares  & \parbox[c]{1em}{\includegraphics[height=0.8in]{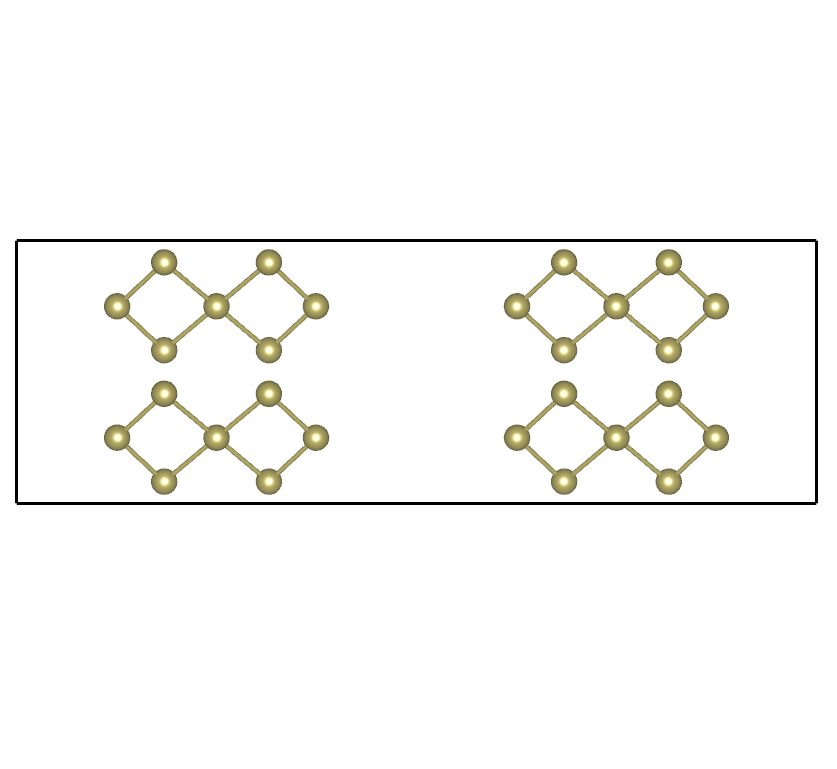}} \\\hline
      3.Linear chain &  \makecell{90[5]: \ch{NbS3 [Nb]}} & & \parbox[c]{1em}{\includegraphics[height=0.8in]{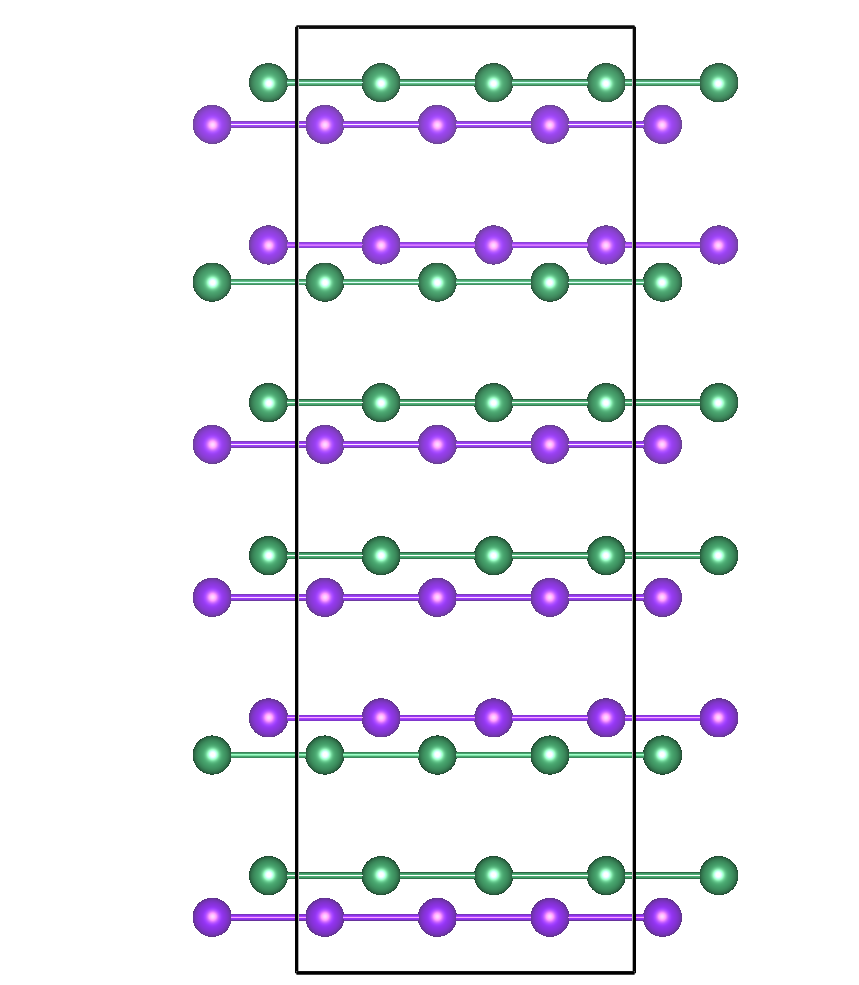}} \\ \hline
      4.Isolated quadrilaterals & \makecell{41[7]: \ch{MoSe3 [Mo]}} & & \parbox[c]{1em}{\includegraphics[height=0.8in]{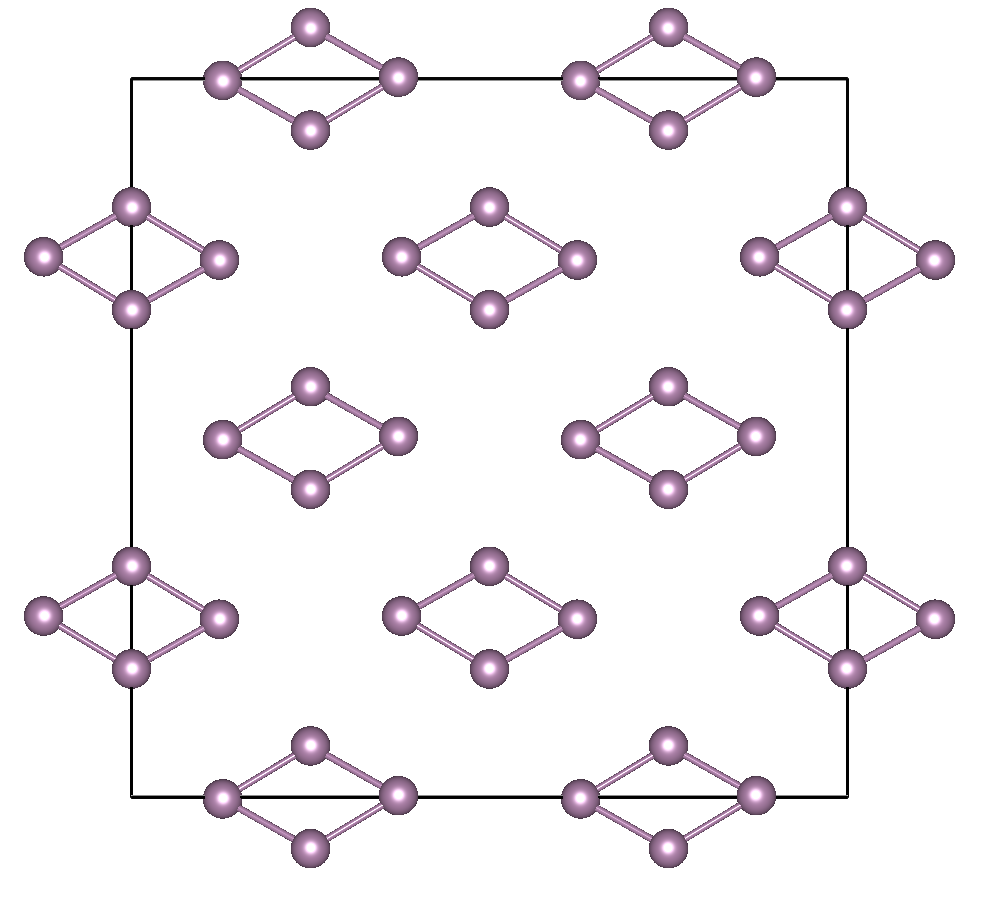}} \\ \hline
      5.$(3^2,4,3,4)$ Archimedean lattice  &  \makecell{38[10]:\ch{TiP [P]}}  &    & \parbox[c]{1em}{\includegraphics[height=0.6in]{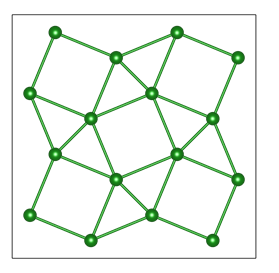}} \\ \hline
      \label{mono}
\end{longtable*}

\begin{longtable*} {|p{0.15\textwidth}|p{0.36\textwidth}|m{0.2\textwidth}|p{0.24\textwidth}|}
  
    \caption{New bilayer flatband sublattices  identified from Fig \ref{fig6}} \\ \hline
    Lattice type & Cluster ID\# [cluster size]:Example [sublattice] & Remarks & Figure \\\hline
    
    1.Stacked Kagome lattice (AA stacking)  &  \makecell{6[4]: \ch{TaIr3[Ir]}}  &  Two stacked Kagome lattice (each has out of plane distortions)  & \parbox[c]{1em}{\includegraphics[height=0.8in]{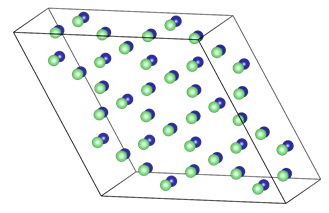}} \\\hline

    2.Stacked centered orthorhombic lattice  &  \makecell{23[4]:\ch{SrLaBr5[Br]}, \ch{BaLaCl5[Cl]}}  &  Elongated hexagonal lattices stacked such that each atom has 6 fold in plane coordination and 2 fold out of plane coordination  & \parbox[c]{1em}{\includegraphics[height=0.8in]{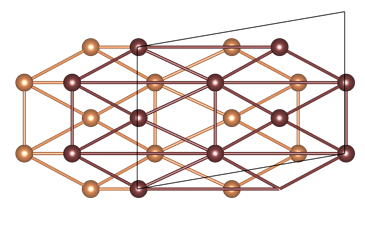}} \\\hline
    3.Stacked rectangles (AB stacking) lattice  &  \makecell{52[7]: \ch{Ba2P [P]}}  &  Two rectangle lattices stacked with AB stacking.   & \parbox[c]{1em}{\includegraphics[height=0.8in]{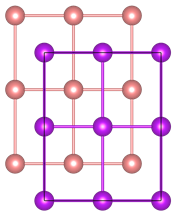}} \\\hline
    4.Stacked hexagonal (AA stacking)  &  \makecell{59[5]:\ch{TiO2[O]} \\ 118[10]: \ch{Rb3Nb2Br9 [Nb]}, \ch{CuAsO3 [Cu]}}  &  Only one nearest neighbour because of stacking  & \parbox[c]{1em}{\includegraphics[width=1.2in]{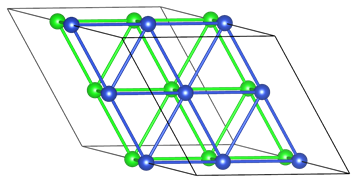}} \\\hline
    5.Linked sawtooth lattice & \makecell{88[9]: \ch{TaI4 [Ta]} } & 2-fold OOP and 2-fold IP coordination & \parbox[c]{1em}{\includegraphics[width=0.8in]{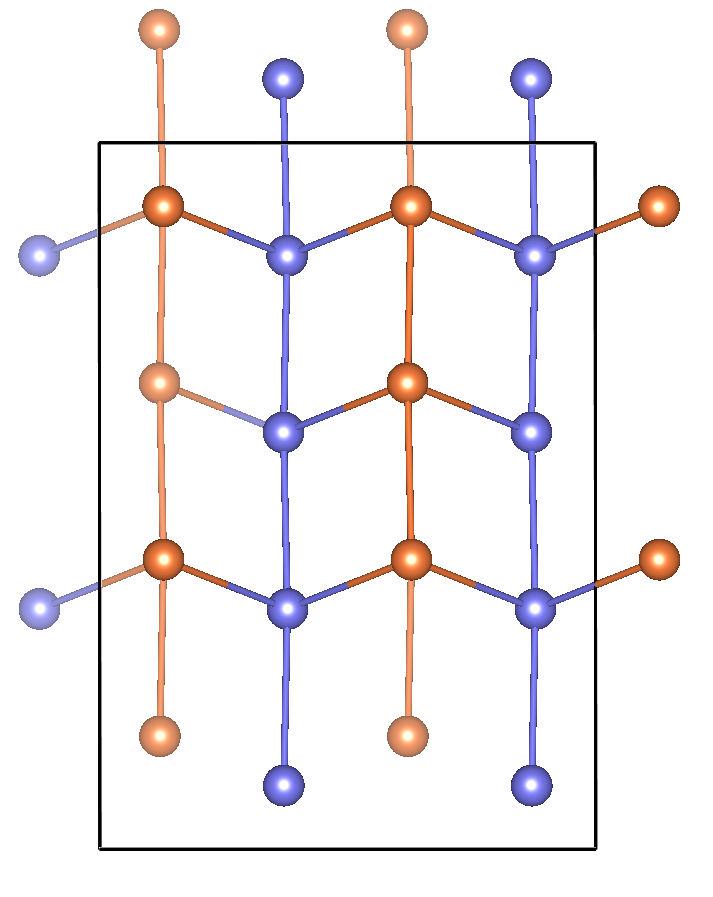}} \\\hline
    6.Stacked centered-orthorhombic-square chain &  \makecell{115[6]:\ch{SmF3 [F]} \\ 121[8]: \ch{CaThBr6 [Br]},  \ch{TbCl3 [Cl]}}  &   Type of Archimedean lattice: Lattice consists of alternate chains of cenetered orthorhombic and square shapes & \parbox[c]{1em}{\includegraphics[height=0.8in]{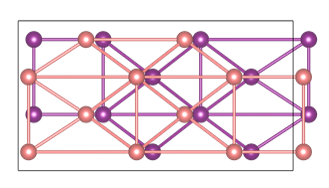}} \\\hline
    7.Reverse sawtooth lattice & \makecell{119[4]: \ch{TaCoTe2 [Co]}, \ch{LiVF5 [V]} \\ 125 [16]: \ch{PdCl2 [Pd]}} & & \parbox[c]{1em}{\includegraphics[width=0.8in]{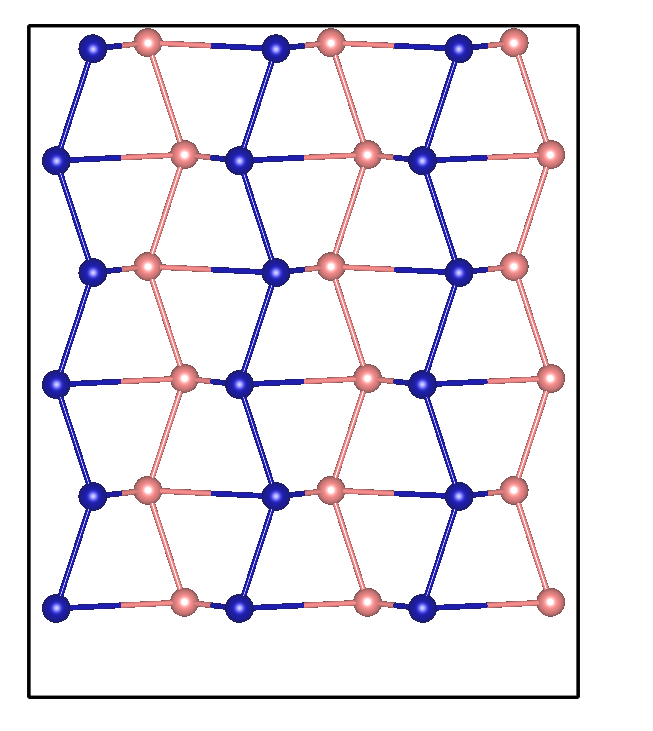} } \\\hline
    \label{bilayer}
\end{longtable*}

\begin{longtable*}  {|p{0.15\textwidth}|p{0.36\textwidth}|p{0.2\textwidth}|p{0.24\textwidth}|}
    \caption{New multilayer flatband sublattices  identified from Fig \ref{fig6}}  
    \\ \hline  
    Lattice type & Cluster ID\# [cluster size]:Example [sublattice] & Remarks & Figure \\\hline

    1.Isolated twisted-octahedra  &  \makecell{15[4]:\ch{ReSeCl[Re]}, \ch{MoCl2[Mo]}}  &  Alternate rows of twisted octahedras  & \parbox[c]{1em}{\includegraphics[height=1in]{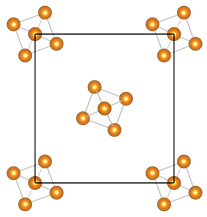}} \\\hline

    2.Stacked square-tetrahedra chain &  \makecell{18[6]:\ch{Sr2TiCu2O7[O]}, \ch{Sr2Cu2SbO7[O]}}  &   & \parbox[c]{1em}{\includegraphics[height=1in]{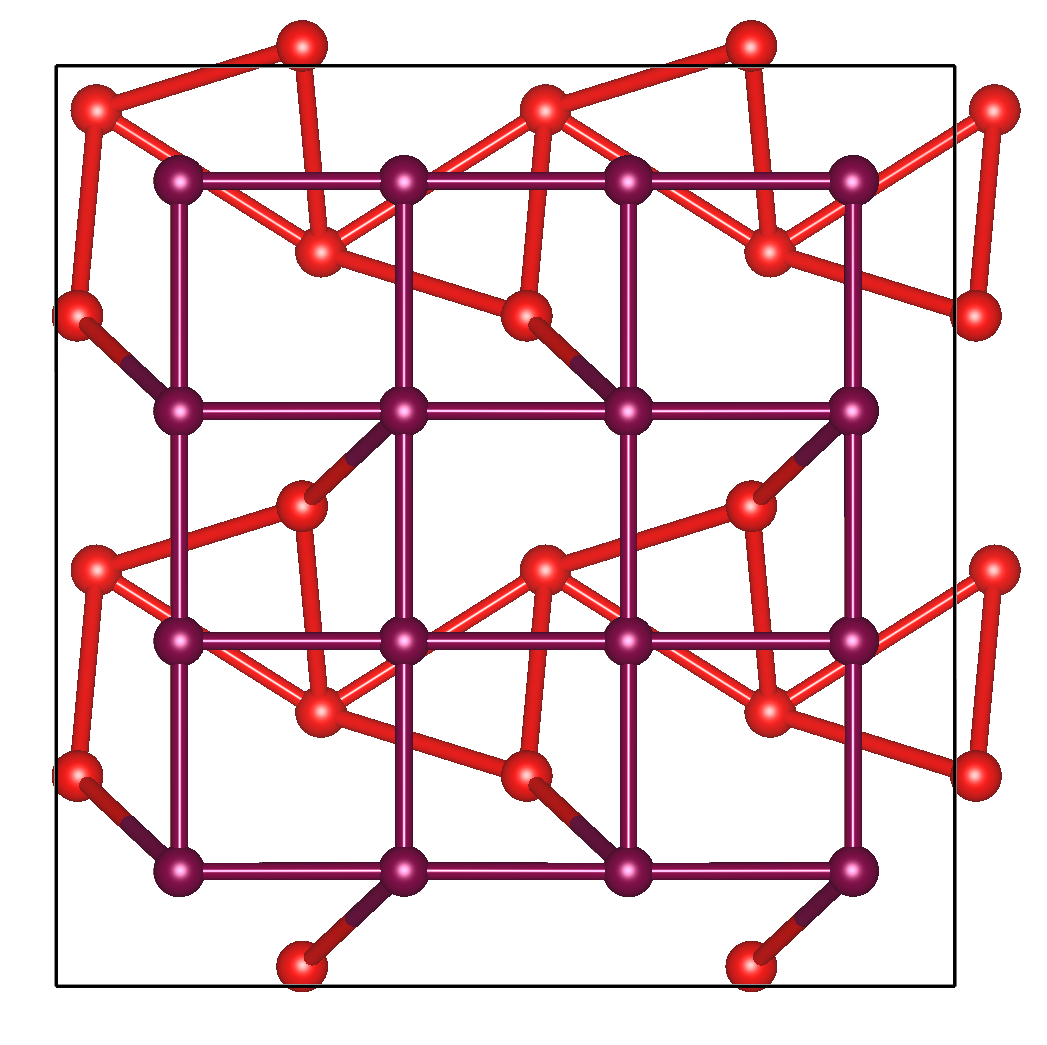}} \\\hline

    3.Distorted stacked hexagonal lattice &  \makecell{25[4]: \ch{BeBr2[Br]} \\ 82[5]: \ch{Fe3S4 [Fe]}}  &  Each atomic layer is hexagonal (monoclinic distortion present)  & \parbox[c]{1em}{\includegraphics[height=1in]{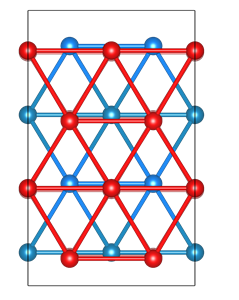}} \\\hline

    4.Stacked colouring triangle-hexagonal lattice  &  \makecell{29[4]:\ch{Li2Fe(PO4)2[O]}, \ch{Ni(ClO4)2[O]}}  &  Two colouring triangle lattices stacked with two hexagonal lattices on ends  & \parbox[c]{1em}{\includegraphics[height=1in]{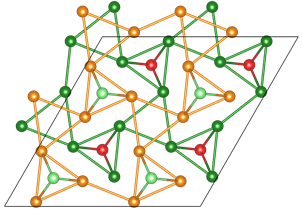}} \\\hline

    5.Stacked $(3^2,4,3,4)$ Archimedean lattice   &  \makecell{91[5]: \ch{CoBr4 [Br]}}  &  2 layers of Archimedean lattice; The Archimedean lattice is distorted out-of-plane  & \parbox[c]{1em}{\includegraphics[height=1in]{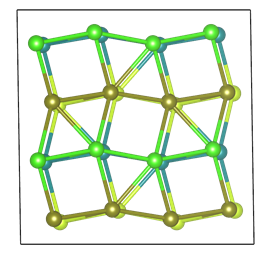}} \\\hline

    6.Vertex-sharing triangular pyramid chain &  \makecell{92[4]:\ch{V2CuO6[O]}, \ch{V2CoO6[O]}}  &    & \parbox[c]{1em}{\includegraphics[height=1in]{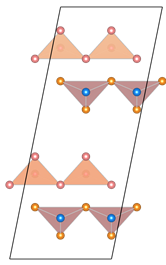}} \\\hline

    7.Connected sawtooth chains & \makecell{94[5]: \ch{In4Se3 [Se]}, \ch{Tl4S3 [S]}} & & \parbox[c]{1em}{\includegraphics[height=1in]{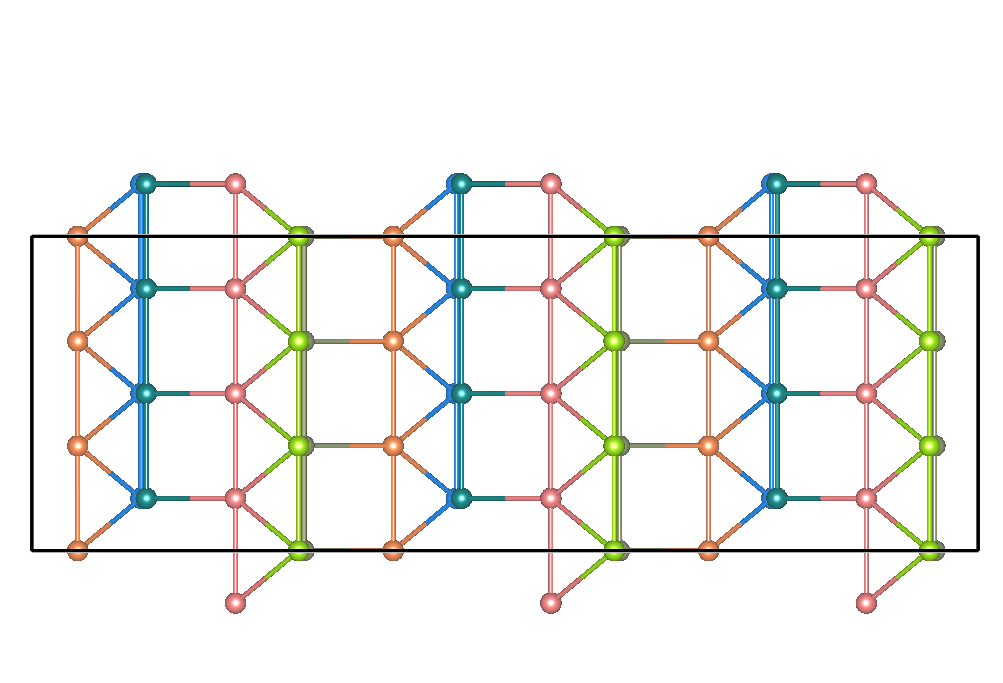}} \\\hline

    8.Edge sharing octahedra chain (zigzag) & \makecell{97[14]:\ch{ZnMoO4[O]}, \ch{ZnWO4[O]}}  &   & \parbox[c]{1em}{\includegraphics[height=1in]{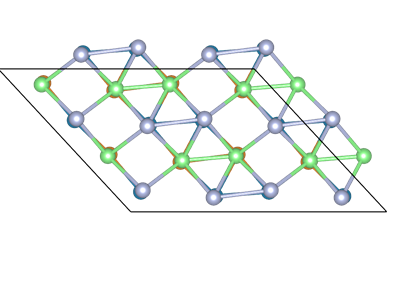}} \\\hline

    9.Vertex-edge sharing planar octahedra &  \makecell{104[4]: \ch{CaTlCl3[Cl]}, \ch{TlPbI3 [I]}}  &    & \parbox[c]{1em}{\includegraphics[height=1in]{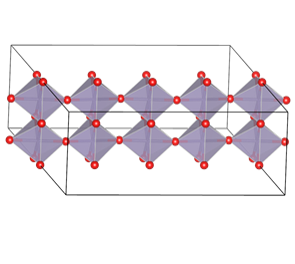}} \\\hline

    \label{multilayer}
\end{longtable*}

\subsection*{Supplementary Note II: Visualization of flatband clusters using UMAP }
\begin{figure}
    \centering
    \includegraphics[width=0.5\textwidth]{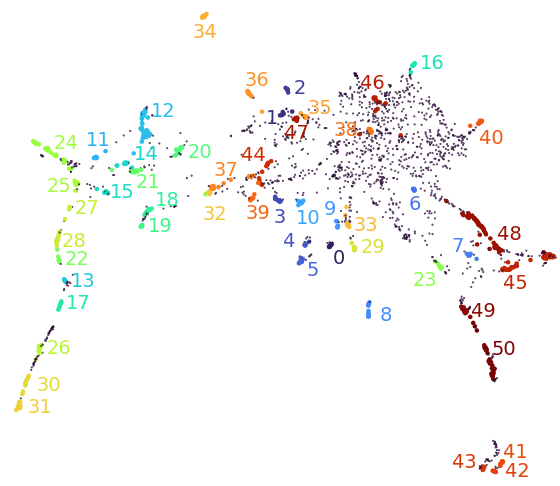}
    \caption{Optimized UMAP visualization (Nearest$\_$neighbor = 30) for HDBSCAN output of MS=7 and SS=6.}
    \label{fig7}
\end{figure}

UMAP \cite{mcinnes2018umap} is another nonlinear dimensionality reduction algorithm, similar to t-SNE, that we have used to present our results. We visualized HDBSCAN output of MS=7, SS=6 using UMAP in 2 dimensions to assess its effectiveness in comparison to that of t-SNE. For UMAP, we optimized the 3 following parameters, nearest$\_$neighbors, minimum$\_$distance, and distance metric. Nearest$\_$neighbors and minimum$\_$distance denote the number of neighbor distances that are retained while reducing the dimensions and minimum distance within a cluster in the embedded space respectively. Here we show clustering results with Nearest$\_$neighbors=30 and minimum$\_$distance=0.1, which is considered to be the best visualization. The resulting embedded space is shown in Fig. \ref{fig7}.Despite the seemingly different presentation, it shares many similar features with that of our t-SNE clustering results. For example, clusters \#41-43 are well isolated from the rest of the clustering. Meanwhile, the transition from hexagonal to rectangular sublattices (mostly yellow and green in color) which have very few outlier points in between, as well as the dispersed distribution of structural fingerprints on the opposite site, are all in good agreement with the t-SNE plot in Fig. \ref{fig5}b. The clustering solution visualized by UMAP method, therefore, does not offer a significantly better solution than the t-SNE method used in the main text.

\bibliography{apssamp}

\end{document}